\documentstyle[aps,12pt,epsf]{revtex}
\begin{document}
\title{Exact solution of the nuclear pairing problem}
\author{Alexander Volya, B. Alex Brown, and Vladimir Zelevinsky}
\address{Department of Physics and Astronomy and \\
National Superconducting Cyclotron Laboratory, \\
Michigan State University,
East Lansing, Michigan 48824-1321, USA}
\maketitle
\begin{abstract}
\baselineskip 14pt
In many applications to finite Fermi-systems, the pairing problem has to be 
treated exactly. We suggest a 
numerical method of exact solution based on SU(2) quasispin
algebras and demonstrate its simplicity and practicality. We show that the 
treatment of binding
energies with the use of the exact pairing and uncorrelated monopole 
contribution of 
other residual interactions can serve as an effective alternative to
the full shell-model diagonalization in spherical nuclei.
A self-consistent combination of the exactly treated pairing and Hartree-Fock
method is discussed. Results for Sn isotopes 
indicate a good agreement with experimental data.
\end{abstract}
%
Pairing correlations play an essential role in nuclear structure properties
including
binding energies, odd-even effects, single-particle occupancies, excitation
spectra,
electromagnetic and beta-decay probabilities,
transfer reaction amplitudes, low-lying collective modes, level densities,
and moments of
inertia \cite{Bel,BM,hasegawa93}. 
The revival of interest in pairing correlations is 
related to studies
of nuclei far from stability and predictions of exotic
pairing modes \cite{Goo,broglia00}.
Metallic clusters, organic molecules and Fullerenes are other examples of 
finite Fermi systems with possibilities for pairing correlations of the
superconducting type \cite{clu}.

The conventional description of pairing usually employs
the classical BCS approach \cite{BCS57} used in 
theory of superconductivity. This
approximate solution has a very good  accuracy for large systems and   
becomes exact in the asymptotic limit \cite{asymp}. The shortcomings of the BCS
approximation for small systems are well known, see for example 
\cite{RingSchuck} and references therein.
The major drawback of the BCS is the violation of particle number
conservation, which gives rise to deviations
from the exact solution for small systems.
Various ideas have been suggested to correct this deficiency, such as the   
number projection 
mean-field methods  \cite{lipkin60,nogami64,satula2000},
coherent state approach \cite{rowe},
stochastic number projection
\cite{capote99}, statistical descriptions
\cite{gupta}, treatments of 
residual parts of the Hamiltonian in the random phase
approximation \cite{bang70,johns70},
and recurrence relation methods \cite{dang66,andreozzi85}. 
These methods have 
found only a limited number of practical
applications; for some approaches the obtained results did not manifest
the desired accuracy whereas
other methods are limited by practical complications.
BCS-like approximate theories have a number of other deficiencies when applied
to small systems \cite{Bel,RingSchuck}. In particular, in the region of weak 
pairing the BCS has a sharp
phase transition from the paired condensate to the normal state with no pairing
(trivial or zero gap solution), whereas exact solutions
exhibit the existence of exponentially decreasing pairing correlations
all the way down to the zero pairing strength.
This difficulty makes the BCS method unreliable for applications to
weakly bound nuclei and calls for improvements and extensions
such as BCS+RPA \cite{hagino2000}. There are also serious problems related to
the correct description of pairing in excited states.

The exact pairing (EP) method presented in this work allows one to
solve exactly the 
general pairing Hamiltonian
\begin{equation}
H=\sum_{j\,m}\epsilon_{j} a^{\dagger}_{j\,m} a_{j\,m} \,+\,
{1\over 4} \sum_{j,\,j^{\prime}} G_{j\,j^{\prime}}\,\sum_{m,\,m^{\prime}}\,
a^{\dagger}_{j\,m} {\tilde a^{\dagger}_{j\,m}}
{\tilde a_{j^{\prime}\,m^{\prime}}}
a_{j^{\prime}\,m^{\prime}
}\,,\quad {\tilde a_{j\,m}}\equiv (-1)^{j-m}a_{j\,-m}\,,
\label{1}
\end{equation}
where $\{\epsilon_j\}$ is the set of single-particle energies, diagonal
$G_{j\,j}$ are pairing energies, and 
$G_{j\,j^{\prime}}=G_{j^{\prime}\,j}$ for $j\ne j^\prime$ are pair
transfer matrix elements, ($j^{\prime}\leftrightarrow j$).
The practical usefulness of the EP algorithm comes from the facts that it is
exact, fast and allows a straightforward extension for an approximate treatment
of other components of
residual interactions. Some realistic examples are presented in this work to
emphasize these points.
As a result, it becomes 
unnecessary to use complex approximate methods, such as BCS and all those
associated with it, when exact results that are free of all
problems discussed above can be obtained
with an almost equal or even smaller effort.

A number of methods for treating the pairing problem exactly have been
previously proposed. 
The Richardson method, described in
the series of papers 
\cite{richardson63_1},
provides a formally exact way for solving the pairing
Hamiltonian. This method reduces the 
large-scale diagonalization of a many-body Hamiltonian in a truncated Hilbert
space to a set of coupled equations with a dimension equal to the 
number of valence particles. Recently, exact solutions 
have been approached by introducing sophisticated mathematical tools 
such as infinite-dimensional algebras \cite{pan98}. Such formally exact
solutions have a certain merit from a mathematical point of view and
for developing and understanding approximate calculations.
However, due to their complexity they are not very useful in solving practical
problems in nuclear physics.

The natural way of solving the pairing problem is related to 
the direct Fock-space diagonalization. For deformed nuclei with the doubly
degenerate single-particle orbitals this approach supplemented by the
appropriate use of symmetries and
truncations was already shown to be quite effective \cite{burglin96,molique97}.
Our goal is to combine the exact treatment of pairing with the approximate
inclusion of other parts of the residual interaction. Here the framework of the
rotationally invariant shell model is the most convenient, especially because
it allows us to fully utilize  
the well known ideas (see for example \cite{kerman61,rowe}),
based on the existence of quasispin symmetry  in paired systems
first studied in the 1940's  by Racah \cite{racah}.
In the context of the shell model similar ideas were utilized in
\cite{auerbach65}.
We use the quasispin
algebra for each subset of
degenerate single-particle levels. The fact that
Racah's degenerate model is analytically
solvable comes purely from this algebraic feature. The spherical
shell model with its {\it m}-degeneracies is a perfect arena for applying 
this method and therefore we will use notations associated with
the case of spherical symmetry and {\it j}-{\it j} coupling.
This certainly does not limit the
generality of approach.

The Hamiltonian Eq. (\ref{1})
can be rewritten as
\begin{equation}
H=\sum_j\, \epsilon_j\Omega_j+2\,\sum_j\, \epsilon_j L^{z}_j+ \sum_{j\,
j^{\prime}}\,G_{j\,j^{\prime}}L^{+}_j L^{-}_{j^{\prime}}\,,
\label{4}
\end{equation}
by introducing the partial quasispin
operators $L^+_j$, $L^-_j$ and $L^z_j$ for each $j$-level as follows
\begin{equation}
L^{-}_j=\frac{1}{2}\,\sum_{m}\, {\tilde a_{j\,m}}
a_{j\,m}\,,\quad
L^{+}_j=\left (L^{-}_j \right)^\dagger=\frac{1}{2}\,\sum_{m}\,
a^{\dagger}_{j\,m} {\tilde a^{\dagger}_{j\,m}}\,,
\end{equation}
\begin{equation}
L^{z}_j=\frac{1}{2}\,\sum_{m}\,
\left (a^{\dagger}_{j\,m} a_{j\,m}-\frac{1}{2} \right )\,=
\frac{1}{2}(N_j-\Omega_j)\,,
\label{3}
\end{equation}
where $N_j$ is the particle number operator and $\Omega_j=(2j+1)/2$ is the pair
degeneracy of a given single-particle level $j\,.$
It can be shown directly from the definitions of
$L^+_J$, $L^-_j$ and $L^z_j$ that they form an SU(2) algebra
of angular momentum,
\begin{equation}
\left[L^{+}_j,L^{-}_{j^{\prime}}\right]=2\delta_{j\,j^{\prime}}\,L^{z}_j\,,\,\,
\left[L^{z}_j,L^{+}_{j^{\prime}}\right]=\delta_{j\,j^{\prime}}\,L^{+}_j\,,\,\,
\left[L^{z}_j,L^{-}_{j^{\prime}}\right]=-\delta_{j\,j^{\prime}}\,L^{-}_j\,.
\label{5}
\end{equation}
Expressed in terms of quasispins, the Hamiltonian in (\ref{4}) makes the
pairing problem equivalent to the problem of interacting spins in a magnetic
field, a generalized form of the Zeeman effect \cite{lane64}.
It is clear that every 
square of the partial quasispin 
${\bf L}_j^2=L^{+}_j L^{-}_j-L^{z}_j+(L^{z}_j)^2$ commutes with the Hamiltonian
making $L_j$, corresponding to the eigenvalue
$L_j(L_j+1)$ of ${\bf L}_j^2$,
a good quantum number in the pairing problem. This brings in the major
simplification of the problem.
The maximum value that $L_j$ can take is $\Omega_j/2\,,$ which happens
for a fully paired subshell, such as,
for example, the completely occupied subshell case where
$N_j=2\Omega_j$,
$L^z_j=L_j=\Omega_j/2\,.$ Lower values of the
quasispin quantum number correspond to the Pauli blocking of a part of
the pair
space $\Omega_j$ by $s_j$ unpaired particles.
This reduces the allowed space to $\Omega_j-s_j\,.$
The number $s_j$ can be called the seniority of a
given  $j$-shell. Being related to the quasispin,
$s_j$ are conserved by the pairing
interaction (\ref{4}).
Introducing the partial seniority $s_j$ and partial occupancy $N_j$ by means of
\begin{equation}
L_j=\frac{1}{2}(\Omega_j-s_j)\,\quad L_j^z=\frac{1}{2}(N_j-\Omega)\,,
\end{equation}
we will use $|s_j,N_j\rangle$
instead of the SU(2)
notation $|L_j,L_j^z\rangle\,.$

The quasispin projections $L^z_j\,,$ or the partial occupancies $N_j\,,$
are not  conserved because of  
the pair transfer term  $G_{j\,j^{\prime}}$ (pair vibration).
The usual constraints on angular momentum  
$0\le|L_j^z|\le L_j$ lead to $s_j\le N_j\le 2\Omega_j-s_j$ and 
$s_j\le\Omega_j$ that
have obvious interpretations. Furthermore, since both $L_j$ and $L_j^z$ must be
simultaneously either integers or half-integers, $s_j$ is of
the same parity 
as $N_j\,,$ confirming that particles are transfered only in pairs.
Finally, there is a given total number of particles in the system
$
N=\sum_j\,N_j\,,
$
and one can introduce the total seniority 
$
s=\sum_j\,s_j\,.
$  
Both quantities are conserved by the Hamiltonian and
must be of the same parity.

For each representation given by the set of quantum numbers $\{s_j\}\,,$
one can
construct basis states  $|\{s_j\},\,\{N_j\}\rangle$ going through all 
permutations of $N$ fermions allowed by above constraints.
Finally, a
Hamiltonian matrix can be constructed in this basis using standard properties
of the angular momentum operators
\begin{equation}
L^{\pm}|L,\,L_z\rangle=\sqrt{(L\mp L_z)(L\pm L_z+1)} \,|L,\,L\pm1\rangle\,.
\end{equation}
Diagonal elements become
\begin{equation}
\langle\{s_j\},\,\{N_j\}|H|\{s_j\},\,
\{N_j\}\rangle=\sum_j \left( \epsilon_j N_j+
\frac{G_{j\,j}}{4}(N_j-s_j)(2\Omega_j-s_j-N_j+2)\right )\,,
\label{diagonal}
\end{equation}
and off-diagonal elements that transfer pairs are
$$
\langle\{s_j\},\,\dots N_j+2,\dots N_{j^\prime}-2,\dots|H|\{s_j\},\,
\dots N_j,\cdots N_{j^\prime},\dots \rangle
$$
\begin{equation}
=
\frac{G_{j\,j^{\prime}}}{4}\sqrt{
(N_{j^{\prime}}-s_{j^{\prime}})(2\Omega_{j^{\prime}}-s_{j^{\prime}}-
N_{j^{\prime}}+2)(2\Omega_j-s_j-N_j)(N_j-s_j+2)}\,.
\end{equation}
Diagonalization of this matrix for each representation,
a given set of partial seniorities $\{s_j\}\,,$ is the final step in 
the solution. The
largest Hamiltonian matrix in partial seniority basis,
which corresponds to the lowest allowed 
total seniority,  has a dimension much lower 
than the total fermionic many-body space. Even
for the set of valence orbits encountered in heavy nuclei
it does not exceed several thousands. 
As seen from the above discussion, the Hamiltonian matrix is
very sparse.

Each state of a non-zero seniority is degenerate since $s_j$
unpaired particles 
are untouched by the Hamiltonian and are free to move within a given
$j$-shell, provided that they all remain unpaired.
The number of these fermionic degrees of freedom for each $j$
is  $C(2\Omega_j,s_j)-C(2\Omega_j,s_j-2)$ where $C(m,n)=m!/(n! (m-n)!)$
is a binomial coefficient. The total degeneracy of each state in the $\{s_j\}$
seniority class is
\begin{equation}
\prod_j\,\left [ C(2\Omega_j,s_j)-C(2\Omega_j,s_j-2) \right ]\,.
\end{equation}
The resulting degenerate states that carry seniority quantum
numbers can be further classified  according to 
other symmetry groups of the Hamiltonian such as angular momentum.
All total seniority zero states have spin zero.
For non-zero seniorities, one first has to determine allowed
angular momenta  for a
given subshell $j$ with $s_j$ unpaired particles and then make all possible
couplings of subshells to a total spin.

\begin{figure}
\begin{center}
\epsfxsize=7.0cm \epsfbox{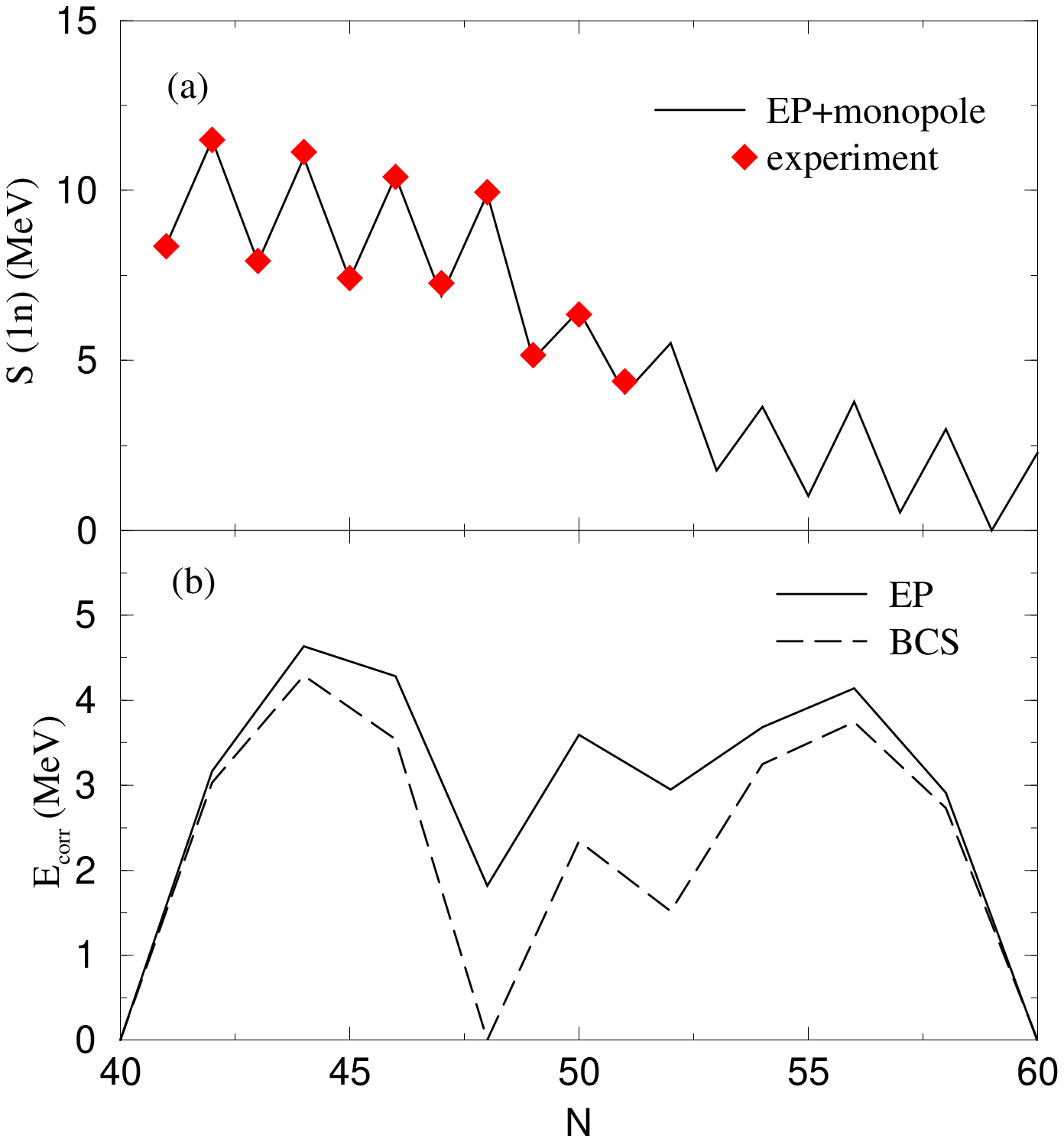}
\end{center}
\caption{Panel (a): neutron separation energies in Ca isotopes,
EP+monopole treatment
of other interactions is compared with experimental points (diamonds);
panel (b): pairing correlation energy in even-even
Ca isotopes, the exact EP calculation
(solid line) is compared with the standard BCS (dashed line).
\label{Ca}}
\end{figure}

To illustrate the practical application of this algorithm, we show below
examples involving chains of Ca and Sn isotopes.
The Ca isotopes occupy the {\it fp} shell
with the model space consisting of four levels $f_{7/2},\,p_{3/2},\,
f_{5/2},$ and $p_{1/2}\,. $ 
This amounts to a total neutron capacity of 20.
The pair transfer matrix elements are taken from the FPD6 interaction
\cite{richter}.
The results in Fig. \ref{Ca}(b) are obtained with single-particle energies
appropriate for the above-mentioned $j$-levels in
$^{48}$Ca: -9.9,  -5.1,  -1.6, and  -3.1 MeV, respectively. 
Fig. \ref{Ca}(b) shows the
correlation
energies in the Ca isotopes, defined as
\begin{equation}
E_{\rm corr}=E-\sum_{j} \epsilon_j \bar{N_j}- \sum_{j}
\frac{G_{j\,j}}{2\Omega_j-1}\,\frac{
\bar{N_j}(\bar{N_j}-1)}{2}\,,
\label{pcorr}
\end{equation}
where $E$ is the energy of the ground state and $\bar{N_j}$ is the (noninteger)
expectation value of the ground state occupancy
of the $j$-level, found as a result of the EP calculation.
It is known \cite{Bel} that in a system with a discrete single-particle
spectrum the nontrivial BCS solution exists only at a sufficiently strong
pairing interaction which can overcome the single-particle level spacings.
In contrast to that, in macroscopic Fermi-systems the nontrivial
Cooper phenomenon exists at any strength of the attractive pairing interaction.
The fact of inadequacy of the BCS approximation for weak pairing was pointed
out by many authors, for example \cite{RingSchuck,molique97}.
Near the $^{48}$Ca shell closure there is a
4.8 MeV gap between
$f_{7/2}$ and higher single-particle orbitals,
and as it can be seen from Fig. \ref{Ca}(b),
BCS can no longer
support the pairing condensate. It gives instead a normal Fermi-gas
solution with zero pairing energy,
whereas in reality the EP solution demonstrates significant pairing effects
with almost 2 MeV
condensation energy.
In most of the real cases the pure pairing interaction ($J=0,\,T=1\,$)
is strongly modified by  residual interactions in other channels. A significant
part of the remaining correlations in spherical nuclei can be treated 
by
including only monopole (diagonal) part
\begin{equation}
\sum_j \bar{V}_{j\,j} \frac{N_j(N_j-1)}{2}+\sum_{j\ne
j^{\prime}}
\bar{V}_{j\,j^{\prime}} N_j N_{j^{\prime}}\,,
\end{equation}
where
\begin{equation}
\bar{V}_{j\, j^{\prime}}=\frac{1+\delta_{j\,
j^{\prime}}}{2\Omega_j (2\Omega_{j^{\prime}}-\delta_{j\,
j^{\prime}})}
\sum_{J\ne 0} (2J+1) \langle J;\,j\,j^\prime|V|J;\,j\,j^\prime\rangle\,   
                                                      \label{mono}
\end{equation}
is the monopole-monopole part of the total interaction $V$ written in eq. 
(\ref{mono}) in terms of the diagonal matrix elements for the pairs
$[jj']_{J}$.
Pairing $\langle J=0;\,j\,j|V|J=0;\,j^\prime\,j^\prime\rangle=
G_{j\,j^{\prime}}\sqrt{\Omega_j \Omega_{j^{\prime}}}\,$ is not included
in the sum since it is treated exactly by EP and its monopole term is present
in Eq. (\ref{diagonal}).
One-neutron
separation energies shown in Fig. \ref{Ca}(a) were calculated with this
EP plus monopole method, using the FPD6 interaction and single-particle
energies for $^{41}$Ca. These results agree very well with experimentally
observed
separation energies, shown on the figure with diamonds.
Ground state energies, obtained with this method
for Ca isotopes differ from those from the exact 
shell model diagonalization, using the total FPD6 \cite{richter}
interaction,
by less than 0.5 MeV; certainly the EP plus monopole results
are exact for full and empty shells and for one-particle or one-hole
cases. 

\begin{figure}
\begin{center}
\epsfxsize=9.0cm \epsfbox{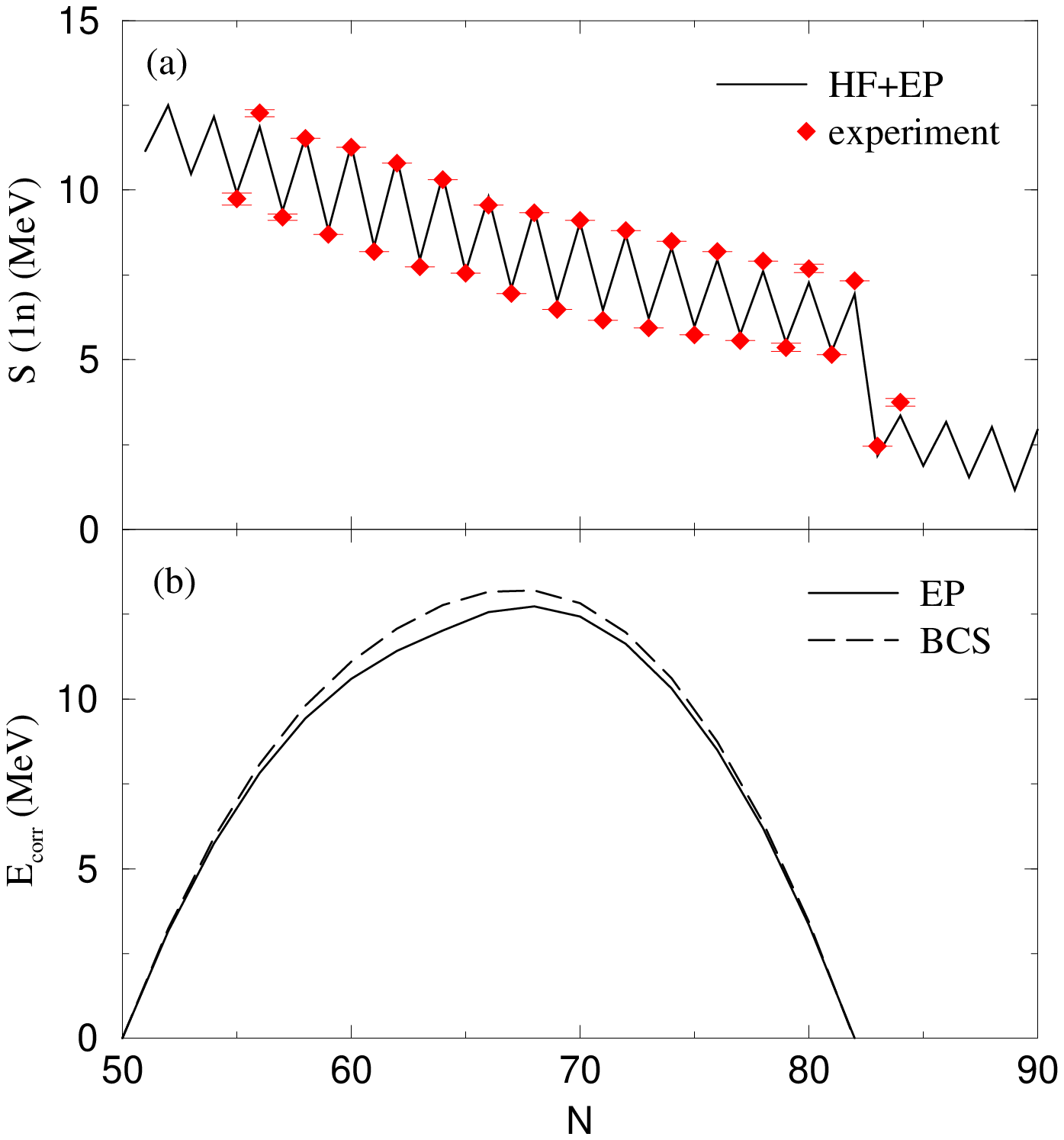}
\end{center}
\caption{Panel (a): neutron separation energies calculated for the Sn isotopes
using the self-consistent solution of Hartree-Fock plus Exact Pairing.
Experimental points are shown for comparison. Panel (b): pairing correlation
energy in even-even Sn isotopes, EP (solid line) and BCS (dashed line).
\label{Sn}}
\end{figure}

For the second example of application we show in Fig. \ref{Sn}(b) the 
results for the Sn isotopes. The model space here consists
of five single-particle levels
$h_{11/2},\,d_{3/2},\,s_{1/2},\,g_{7/2},\,$ and $d_{5/2}\,,$
which can accommodate up to 32 neutrons. We will be interested in the
evolution of neutron separation and correlation energies along the chain of the
isotpes rather than in the specific spectra which were studied earlier
with a number of approximate methods, for example \cite{andreozzi96},
and can be considered in the future with our exact approach. The pairing
matrix elements were obtained from the
$G$-matrix derived from the recent CD-Bonn \cite{machleidt96}
nucleon-nucleon interaction with
$^{132}$Sn as a closed shell, where the $\hat{Q}$-box method includes all
non-folded diagrams up to the third order in the interaction and sums up the
folded diagrams to infinite order \cite{hjorth95}.
The most complex case is $^{116}$Sn (half-filled shell)
with 601,080,390 many-body states of which 272,828 are of spin zero,
so that even with the use of
the angular momentum projection the problem remains difficult for
direct diagonalization. There are only 420 independent seniority sets in the
seniority basis. 
The largest matrix, $s=0\,,$ has a dimension of $110\,,$
the diagonalization of which is a trivial problem. Solution of the pairing
problem for $^{116}$Sn with the algorithm discussed above is therefore
simple and extremely fast.

The correlation energies obtained with the above $G$-matrix and
single-hole energies
-9.76, -8.98, -7.33, -7.66, and  -7.57 MeV,
based on the data for the $^{131}$Sn isotope \cite{TA96}, are shown in
Fig. \ref{Sn}(b).
A reduction of pairing also 
happens between
$g_{7/2},\,d_{5/2}$
and the rest of the single-particle orbitals $h_{11/2},\,d_{3/2},\,s_{1/2}\,,$
but here BCS is just weakened 
which results only in relatively small
suppression of correlation energy.
A related application of EP is shown in Fig. \ref{Sn}(a)
where one-neutron separation energies are calculated
for the Sn isotopes, including those beyond $^{132}$Sn.
These energies were obtained from the fully self-consistent spherically
symmetric
solution of Hartree-Fock (HF) equations, using the SKX interaction 
\cite{brown98}, with the EP solution based on the above $G$-matrix at
each HF iteration.
The HF+EP method works as follows. An initial guess for the HF
potential is used to obtain a set of spherical single-particle energies and
densities for all of the occupied and valence orbitals. The
single-particle enegies for the valence space plus a fixed set of
two-body matrix elements are used in the EP method to obtain the pairing
correlation energy and the valence single-particle occupation numbers.
The initial set of single-particle densities plus the pairing occupation
factors are used to obtain a new potential with the Skyrme hamiltonian.
This proceedure is iterated until convergence (typically 60 iterations).
The total energy is the sum of the Skyrme HF energy and the pairing
correlation energy Eq. (\ref{pcorr}).   
The plot \ref{Sn}(a) exhibits an odd-even staggering
that can only be attributed to pairing. The agreement between experiment and
theory is excellent given that no parameters have been adjusted for these
particular data. 

The EP algorithm
presented here in our view has a good future as a tool for
different calculations
related to pairing. It is exact, fast and reliable which makes it perfect for
pure shell model calculations with pairing,
fast estimates of binding energies and spectroscopic factors,
for the use as a basis for 
treatment of other interactions, iterative HF calculations and
many other tasks.\\ 
\\
This work was supported by the National Science Foundation,
grants No. 9605207 and 0070911. The schematic model was studied with E. Keller
in the framework of the REU program at MSU \cite{keller2000}. 
The authors thank S. Pratt
for motivating discussions and G. Bertsch, A. Klein,
M. Mostovoy, and D. Mulhall for useful comments. M. Hjorth-Jensen kindly
provided the renormalized $G$-matrix for the $^{132}$Sn region.

\end{document}